\documentclass[conference]{IEEEtran}
\usepackage{tabularx,booktabs,makecell}
\usepackage{algorithm}
\usepackage{algorithmic}
\usepackage{epsfig,graphics,epigraph,graphicx,subcaption, color,xcolor}
\begin{document}

\title{TrafficNet: An Open Naturalistic Driving Scenario Library}

\author{\IEEEauthorblockN{Ding Zhao}
\IEEEauthorblockA{Department of Mechanical Engineering\\
Robotics Institute\\
University of Michigan\\
Ann Arbor, USA\\
zhaoding@umich.edu}
\and
\IEEEauthorblockN{Yaohui Guo}
\IEEEauthorblockA{Robotics Institute\\
University of Michigan\\
Ann Arbor, USA\\
Email: yaohuig@umich.edu}
\and
\IEEEauthorblockN{Yunhan Jack Jia}
\IEEEauthorblockA{Department of Electrical Engineering\\
and Computer Science\\
University of Michigan\\
Ann Arbor, USA\\
Email: jackjia@umich.edu}}
\maketitle

\begin{abstract}
The enormous efforts spent on collecting naturalistic driving data in the recent years has resulted in an expansion of publicly available traffic datasets, which has the potential to assist the development of the self-driving vehicles. However, we found that many of the attempts to utilize these datasets have failed in practice due to a lack of usability concern from the organizations that host these collected data. For example,  extracting data associated with certain critical conditions from naturalistic driving data organized in chronological order may not be convenient for a vehicle engineer that doesn't have big data analytics experiences.

To address the general usability challenges of these publicly available traffic datasets, we propose TrafficNet, a large-scale and extensible library of naturalistic driving scenarios, aiming at bridging the gap between research datasets and practically usable information for vehicle engineers and researchers. The proposed web-based driving scenario database preprocesses massive raw traffic data collected in chronological order into an organized scenario-based dataset by applying a set of categorization algorithms to label the naturalistic driving data with six different critical driving scenarios. TrafficNet opens not only the scenario library but also the source code of these categorization methods to the public, which will foster more sophisticated and accurate scenario-based categorization algorithms to advance the intelligent transportation research. The source code and the scenario database can be accessed at https://github.com/TrafficNet.
\end{abstract}
\IEEEpeerreviewmaketitle

\section{Introduction}

Understanding driving environment is critical for the development of intelligent transportation systems. For instance, a self-driving car is expected to be cognitive of the varying environment, including weather, road conditions, and interactions with other vehicles, pedestrians, and cyclists. The current understanding of the complexity of the traffic scenarios primarily comes from two data sources: the crash database, and the naturalistic driving data.

The crash databases have been established by many countries to collect and analyze the scenarios that may lead to critical safety issues. For instance, the CSD and GIDAS crash databases help to anticipate the benefits of one scenario in terms of the safety, and by analyzing these crash scenarios, researchers can identify and collect a set of key scenarios to build their testing benchmarks. However, due to the limited environmental information available in the crash databases for each accident, it is usually difficult to reconstruct the complete and realistic context of the crash site.

Naturalistic driving data, on the other hand, provides not only the vehicle dynamics information, but also highly detailed data collected from many types of on-board equipments, such as Mobileye~\cite{mobileye}, radar, etc, which represents how people drive in the real world accurately. Additionally, the persistently logged naturalistic driving trace can provide historical data prior to a certain critical event, with which the root cause can be identified. More importantly, as the growing interest in the development of self-driving functionality, the detailed naturalistic driving data collected from various sensors, which will be equipped with all the autonomous vehicles in the future are becoming even more critical.

In a Naturalistic-Field Operational Test (N-FOT), data is collected from a number of equipped vehicles driven in naturalistic conditions over an extended period of time~\cite{aust2012evaluation}. Some of the large-scale N-FOT projects conducted in the U.S. are shown in table~\ref{Table_NFOT}. For example, the 100-Car Naturalistic Driving Study~\cite{neale2005overview} conducted by the Virginia Polytechnic Institute and State University studies the main factors causing vehicle crashes. Its data has been used to analyze driver performance, behavior, environment, driving context and other factors that were associated with critical incidents. While more recently, the Safety Pilot Model Deployment (SPMD)~\cite{spmd} launched by the University of Michigan Transportation Research Institute (UMTRI) was a comprehensive data collection effort under real-world conditions, with multimodal traffic and vehicles equipped with vehicle-to-vehicle (V2V) and vehicle-to-infrastructure (V2I) communication devices. The deployment included approximately 2,800 equipped vehicles and 30 roadside equipment. The data was logged and available in text format.

\begin{table*}[t]
	\centering
	\caption{Existing Traffic Scenarios}
	\label{Table_NFOT}
    \begin{tabular}{l@{\hskip 0.1in}l@{\hskip 0.1in}l@{\hskip 0.1in}l@{\hskip 0.1in}l@{\hskip 0.1in}l@{\hskip 0.1in}l}
     \hline
     \toprule
    Name& Conductor & Period& Mileage[mile]& Vehicle& Sensor  & Drivers\\
    \midrule
     \makecell[l]{100 Car Naturalistic\\ Driving Study \cite{neale2005overview}} & VT        & \makecell[l]{2001-\\2009}    & 2 million & 100 sedans& Camera& \makecell[l]{109 primary   drivers\\132 secondary drivers} \\
     ACAS \cite{ervin2005automotive}     & UM   & \makecell[l]{2004-\\2005}    & 137,000       & 11 sedans     & \makecell[l]{Camera\\Radar}   & 96 drivers\\
     RDCW \cite{leblanc2006road}     & UM   & \makecell[l]{2005-\\2006}    & 83,000        & 11 sedans     & \makecell[l]{Camera\\Radar}   & 11 drivers\\
     SeMiFOT \cite{victor2010sweden}       & UM   & \makecell[l]{2008-\\2009}    & 106,528       & \makecell[l]{10 sedans,\\4 trucks}    & \makecell[l]{Camera\\Radars}  & 39 drivers      \\
     IVBSS \cite{sayer2011integrated,leblanc2011driver}    & UM   & \makecell[l]{2010-\\2011}    & \makecell[l]{sedan: 213,309\\   truck: 601,944} & \makecell[l]{16 sedans\\10 heavy trucks}   & \makecell[l]{Camera\\Radar}   & \makecell[l]{108 drivers for sedans  \\18 professional truck drivers}   \\
     SPMD \cite{sayer2011integrated}     & UM   & \makecell[l]{2012-\\present}    & Over 34 million    & \makecell[l]{2,800 various\\   types of vehicles} & \makecell[l]{Camera \\DSRC}     & \makecell[l]{2,700 volunteer drivers \\ several professional bus and truck drivers} \\
     \makecell[l]{Google driverless\\car \cite{googlereport}}    & Google    & \makecell[l]{2012-\\present} & 2 million   & \makecell[l]{At least 50\\sedans and SUVs}     & \makecell[l]{Lidar\\Camera \\ Radar} & Google   technicians and volunteers      \\
     SHRPII \cite{punzo2011assessment}   & US DOT    & \makecell[l]{2006-\\2015}       & 35 million        & 3,500+ vehicles        & \makecell[l]{Camera\\Radar} & 3,500+ drivers     \\
    \bottomrule
     \hline
\end{tabular}
\end{table*}

However, all the traffic data are logged and organized in the chronological order, and requires extensive post-processing to extract useful information such as the scenarios that the users are interested in. This heavy and tedious data processing and data mining work usually discourage the potential dataset users, and thus downgrade the value of these data collected with huge efforts. Moreover, these datasets contain the raw sensor data, collected from on-board equipment such as radar, Mobileye, which usually results in extremely large file size, and thus raises the bar for vehicle engineer and researchers without the big data analytics background, or sufficient computing resources to utilize them. We believe that a well-organized and maintained scenarios library will address the usability issue, and significantly increase the ITS development efficiency.

TrafficNet, a web-based library for driving scenarios is designed to reduce efforts for to use the open datasets, and thus bridge the gap between research datasets and practically usable information for vehicle engineers and researchers. The data currently used in TrafficNet to construct the scenario library is from the U.S. Department of Transportation (DoT) Data Acquisition System (DAS1), which is collected through the SPMD project conducted by the University of Michigan. The schema of the TrafficNet microscopic library is described in Section II, while the construction of the six key driving scenarios is described with statistical analysis in Section III. We are working to include more types of scenarios and plan to extend our analysis to other open database as well. In the current version, although with our best efforts, the database may still contain wrongly labeled events. We will improve the accuracy of the database in the future version using automated and manual approaches. Meanwhile, we open our source code and allow remote users to submit their own queries to potentially accelerate the development by leveraging on crowdsourcing. The source code and the scenario database can be accessed at https://github.com/TrafficNet.

\section{Construction of the TrafficNet}
TrafficNet is an ambitious project. Thus far, we have constructed the scenario datasets for six major real-world driving scenarios, which are among the most widely-used scenarios in the vehicle evaluation, and the testing of self-driving functionalities. TrafficNet uses the relational database to store and organize the large volumes of traffic data from the open repository of DoT. The raw data mainly consist of five different types of events listed below.

\begin{itemize}
\item \textbf{Front targets} data elements are populated with the aid of Mobileye's vision-based Advanced Driver Assistance Systems. It contains information such as the position and velocity of targets and obstacles in front of the vehicle.
\item \textbf{Data lane} is the log of lane-based information collected by the on-board Mobileye sensor that communicates the vehicle's position relative to the lane boundaries.
\item \textbf{Wireless safety unit} data are collected from the on-board WSU, and contains GPS-based data and those that are obtained from the vehicle's Control Area Network (CAN) bus.
\item \textbf{Trip summary} data contains trip-level summaries from each instrument vehicle, for each trip taken during the selected time period of the model deployment. It includes details such as distance traveled, the number of times the driver applied the brake, and the distance with the vehicle speed over 25 mph.
\item \textbf{Radar} data are collected from a radar unit that is part of a vehicle's integrated safety device unit. It contains the estimation of the type of object that is in front of the vehicle, as well as the object's speed and relative location.
\end{itemize}

\begin{figure}[h]
\centering
\includegraphics[width=\columnwidth]{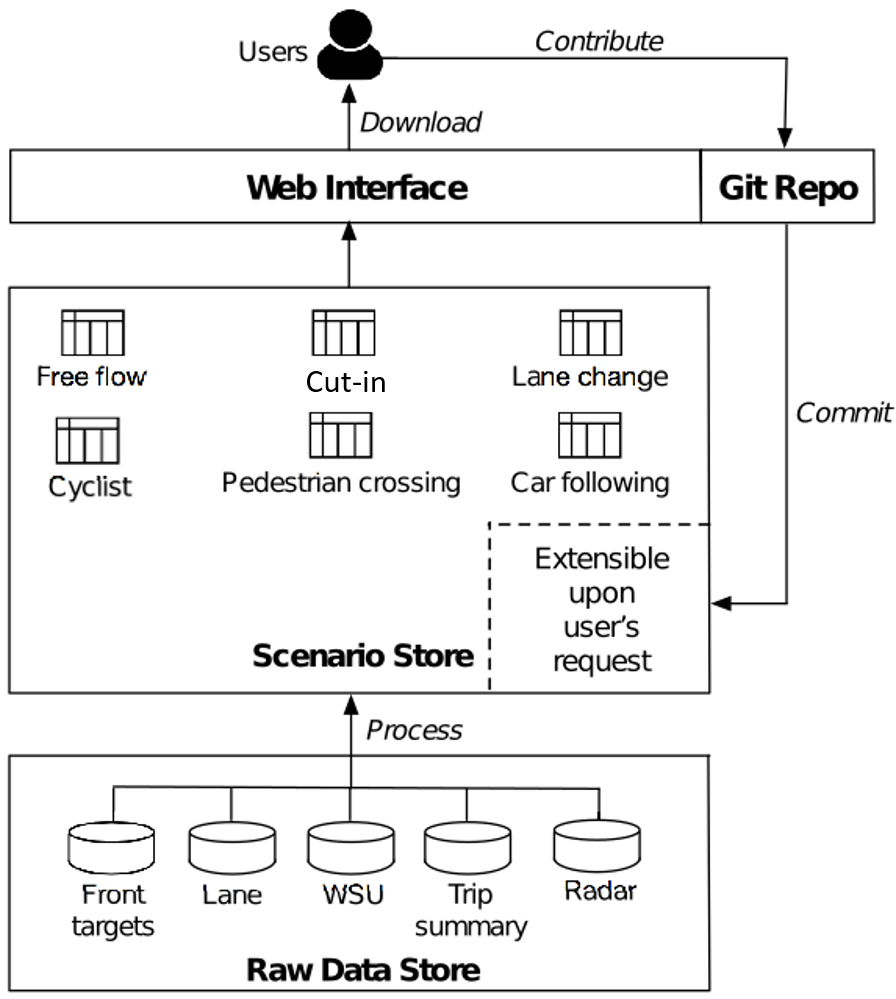}
\caption{TrafficNet] Overview}
\label{fig:overview}
\end{figure}

Shown in Figure~\ref{fig:overview}, the tens of gigabytes of raw data collected in the chronological order are stored in the database of our back-end server, while the processing procedure extracts the data corresponding to each of the six scenarios to build the scenario database.

Compared to the raw data available only in the chronological order, researchers and engineers in the automobile industry usually care more about the real-world scenarios each data record is associated with, so that they can use them to test their vehicles or algorithms against different physical environments. Such information is becoming even more valuable than before, since advanced supervised learning techniques are widely used in the development of self-driving functionalities, whose accuracy highly depends on the correct labeling of these data for their corresponding scenarios.

TrafficNet integrates the extraction of 6 critical scenarios in the current stage, including free flow, car-following, 
lane change, frontal cut-in  
pedestrian crossing,  
and cyclist. The extraction methodology is designed to leverage the information from
raw data tables, and each MySQL~\cite{mysql} script developed by us is used to construct the table for one traffic scenario. Each table contains the data from either the vehicle or the Road Side Unit (RSU) that are corresponding to one specific type of the real-world scenario. For example, the car-following scenario table consists of all the data when an instrumented vehicle was performing the car-following action. All the data in the scenario store are hosted on our web server and can be accessed and downloaded through our web page without limits.


\section{Six Scenario}
This section describes the database we build for six scenarios, as well as the algorithms we used to query them from the SPMD database. There are two tables available for each scenario, namely the $Event$ table and the $Squence$ table. Here we define an event as an individual instance happens in a continuous time interval. The $Event$ table records the primary keys of all the events such as the $Device$, $Trip$, $StartTime$ and $EndTime$ data, while the $Sequence$ table records all the data for each scenario in time sequence. The tables we use here consists of $DataWsu$, $DataFrontTargets$ and $DataLane$. $DataFrontTargets$ and $DataLane$ are collected by Mobileye sensor on target type, position and speed information about the obstacle, and distance to the lanes respectively; $DataWsu$ is recorded by the Wireless Safe Unit (WSU) on the status about the vehicle. A summary of the total number of events in every scenario is shown in Table~\ref{tab:summarytable}.

\begin{table}[]
\centering
\caption{Scenario summary}
\label{tab:summarytable}
\begin{tabular}{ll}
\hline
Scenario            & Total Events \\ \hline
Free Flow           &  440,001             \\
Pedestrain Crossing &  26,412            \\
Cyclist             &  1,270            \\
car-following       &  104,849            \\
Lane change         &  10,873            \\
Cut In              &  72,886
 \\ \hline
Sum             &  565,291
\\ \hline
\end{tabular}
\end{table}

\begin{figure}[t]
  \begin{subfigure}{.50\columnwidth}
      \centering
      \includegraphics[width=\linewidth]{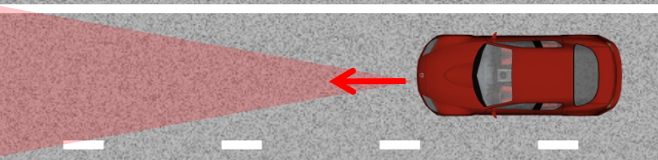}
      \caption{Free flow}
      \label{fig:freeflowa}
  \end{subfigure}
  \begin{subfigure}{.49\columnwidth}
      \centering
      \includegraphics[width=\linewidth]{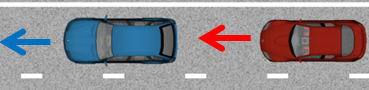}
      \caption{car-following}
      \label{fig:carfollowinga}
  \end{subfigure}
    \begin{subfigure}{.50\columnwidth}
      \centering
      \includegraphics[width=\linewidth]{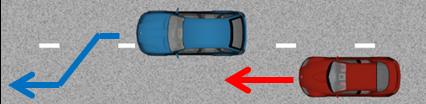}
      \caption{Cut in}
      \label{fig:cutina}
  \end{subfigure}
    \begin{subfigure}{.49\columnwidth}
      \centering
      \includegraphics[width=\linewidth]{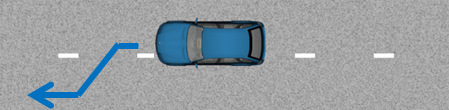}
      \caption{Lane change}
      \label{fig:lanechangea}
  \end{subfigure}
    \begin{subfigure}{.50\columnwidth}
      \centering
      \includegraphics[width=\linewidth]{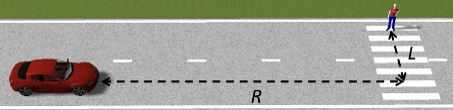}
      \caption{Pedestrian}
      \label{fig:pedestriana}
  \end{subfigure}
    \begin{subfigure}{.49\columnwidth}
      \centering
      \includegraphics[width=\linewidth]{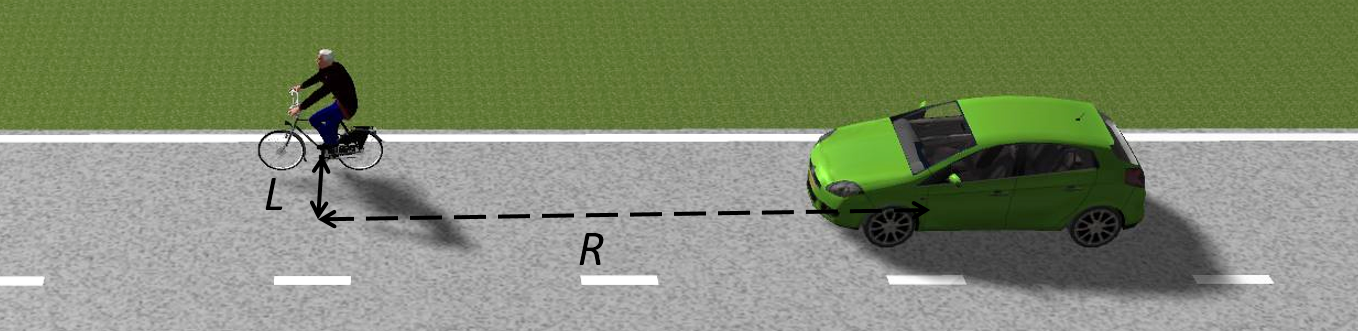}
      \caption{Cyclist}
      \label{fig:cyclist}
  \end{subfigure}

  \caption{Scenarios in TrafficNet}
      \label{fig:scenarios_all}
\end{figure}

\subsection{Free flow}

Free flow is the scenario where a vehicle goes freely without being blocked by other vehicles in front of it~\cite{lee1998origin}. The scenario is illustrated in Fig~\ref{fig:freeflowa}. We use the data of front targets position and GPS information collected by Mobileye and WSU to check whether there are obstacles in front of the host vehicle. If the primary key of $DataWsu$ doesn't appear in $DataFrontTargets$, it indicates there is no front obstacle, so the corresponding row will be selected into the free flow scenario. The algorithm is described in Algorithm~\ref{alg:freeflow}.


 \begin{algorithm}[H]
 \caption{FreeFlow}
 \label{alg:freeflow}
 \begin{algorithmic}[1]
 \renewcommand{\algorithmicrequire}{\textbf{Input:}}
 \renewcommand{\algorithmicensure}{\textbf{Output:}}
 \REQUIRE $DataFrontTargets$, $DataWsu$
 \ENSURE  $FreeFlowEvent$
 \\ \textit{Initialisation} :
 \STATE $FreeFlowEvent$ $\leftarrow \emptyset$ \\
 \textit{Pick rows with no obstacle detected} :
 \FOR {$row$ $\subset$ $DataWsu$,}
 \IF {$row \not\subset DataFrontTargets$}
 \STATE $FreeFlowEvent$$\leftarrow$ $row$
 \ENDIF
 \ENDFOR
 \RETURN $FreeFlowEvent$
 \end{algorithmic}
 \end{algorithm}


\begin{figure}[h]
\centering
\includegraphics[width=\columnwidth]{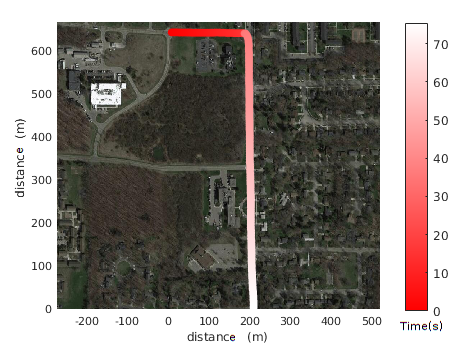}
\caption{Example of free flow events}
\label{fig:freeflowb}
\end{figure}

\begin{figure}[h]
\centering
\includegraphics[width=0.75\columnwidth]{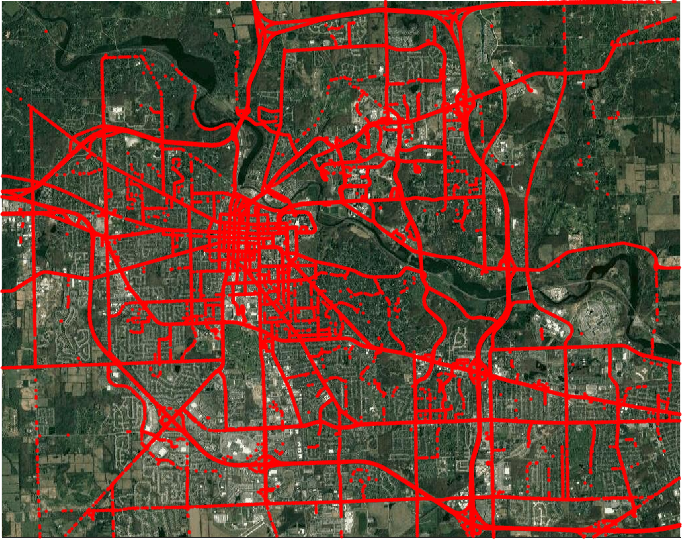}
\caption{Localization of free flow events}
\label{fig:freeflowc}
\end{figure}

We plot each event as a red point in the Google satellite map of Ann Arbor according to the GPS information to visualize the distribution of the events as shown in Fig~\ref{fig:freeflowc}. The free flow events are well distributed in both rural and urban area thus can provide information on the human driver's behavior on different kinds of roads when they can drive freely.


\subsection{Car-following}
car-following contains the events where a vehicle keeps following one car without changing its lane. The dataset contains a column named $fID$ indicates the index of the car-following event. To query this dataset, we exploit the $LaneChangeEvent$ to determine a continuous driving segment $s$ without lane change. Then for each $s$, we divide it into car-following segments at where the $CIPV$ equals 1 in $DataFrontTargets$, which means there exists an obstacle in the path of the vehicle. By doing so, we get individual segments where the vehicle follows the same car in the same lane. The algorithm is illustrated in Algorithm~\ref{alg:carfollowing}. The symbol '$\backslash$' means the relative complement, i.e. a row in $DataFrontTargets\backslash LaneChangeEvent$ means it is in $DataFrontTargets$ while not in $LaneChangeEvent$. This data enables learning the strategy from a human driver on how to keep a safe distance and safe speed when following the front car. An example of using car-following datasets to help develop automated vehicles can be found in \cite{Zhao2016following}.

 \begin{algorithm}[H]
 \caption{car-following}
 \label{alg:carfollowing}
 \begin{algorithmic}[1]
 \renewcommand{\algorithmicrequire}{\textbf{Input:}}
 \renewcommand{\algorithmicensure}{\textbf{Output:}}
 \REQUIRE $DataFrontTargets$, $LaneChangeEvent$
 \ENSURE  $CarFollowing$
 \\ \textit{Initialisation} :
  \STATE $CarFollowing$ $\leftarrow \emptyset$,$fID \leftarrow 0$
  \FOR{each row $r_t$ in \\$DataFrontTargets\backslash LaneChangeEvent$}
  \WHILE{$CIPV_t=1$ and \\$ObstacleId_{t}= ObstacleId_{t-1}$}
  \STATE $row_t \leftarrow fID$
  \STATE $CarFollowing \leftarrow row_t$
  \STATE $t = t + 1$
  \ENDWHILE
  \STATE $fID = fID+1$
  \ENDFOR
 \RETURN $CarFollowing$
 \end{algorithmic}
 \end{algorithm}


\begin{figure}[h]
\centering
\includegraphics[width=\columnwidth]{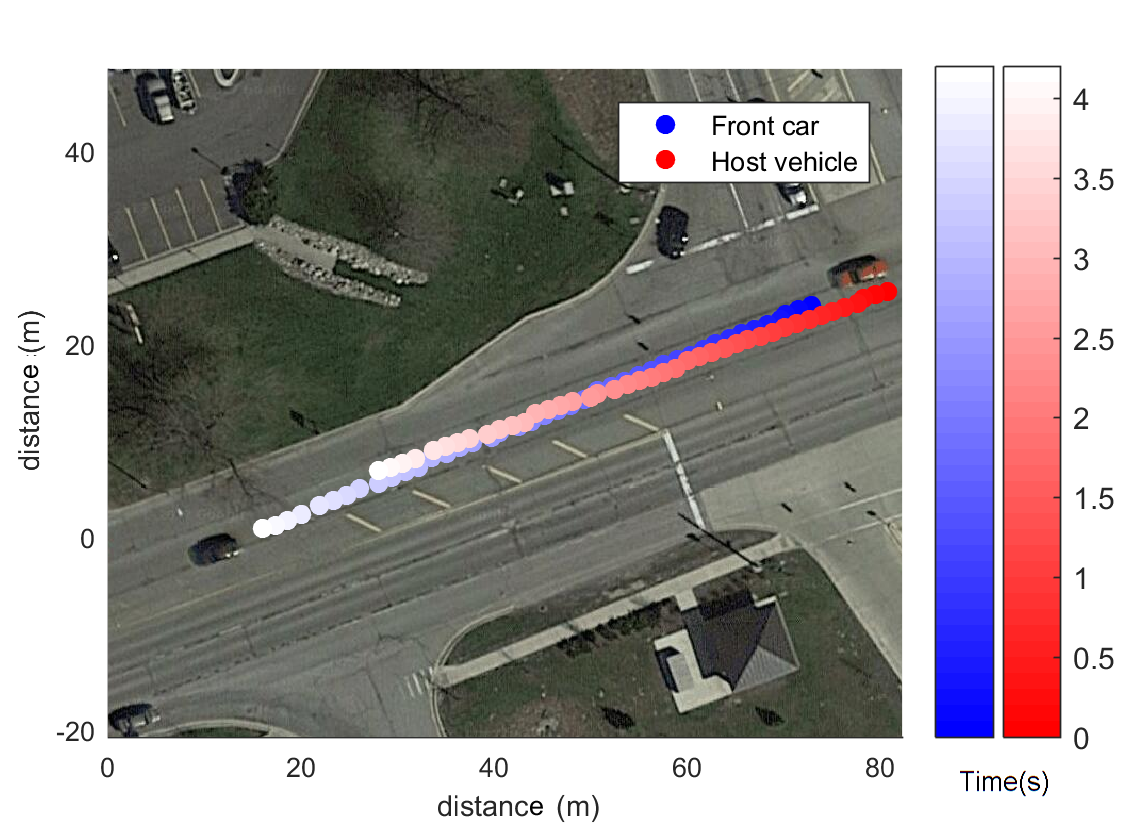}
\caption{Example of car-following events}
\label{fig:carfollowingb}
\end{figure}

\begin{figure}[h]
\centering
\includegraphics[width=0.75\columnwidth]{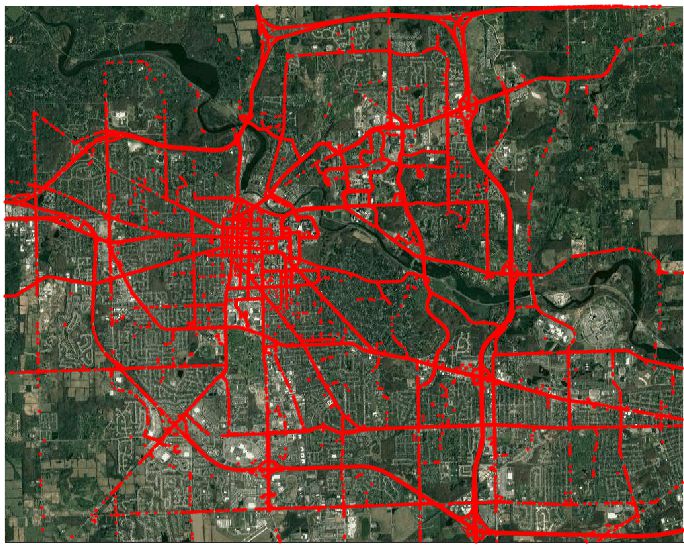}
\caption{Localization of car-following events}
\label{fig:carfollowingc}
\end{figure}




\subsection{Cut In}
\begin{figure}[h]
\centering\includegraphics[width=\columnwidth]{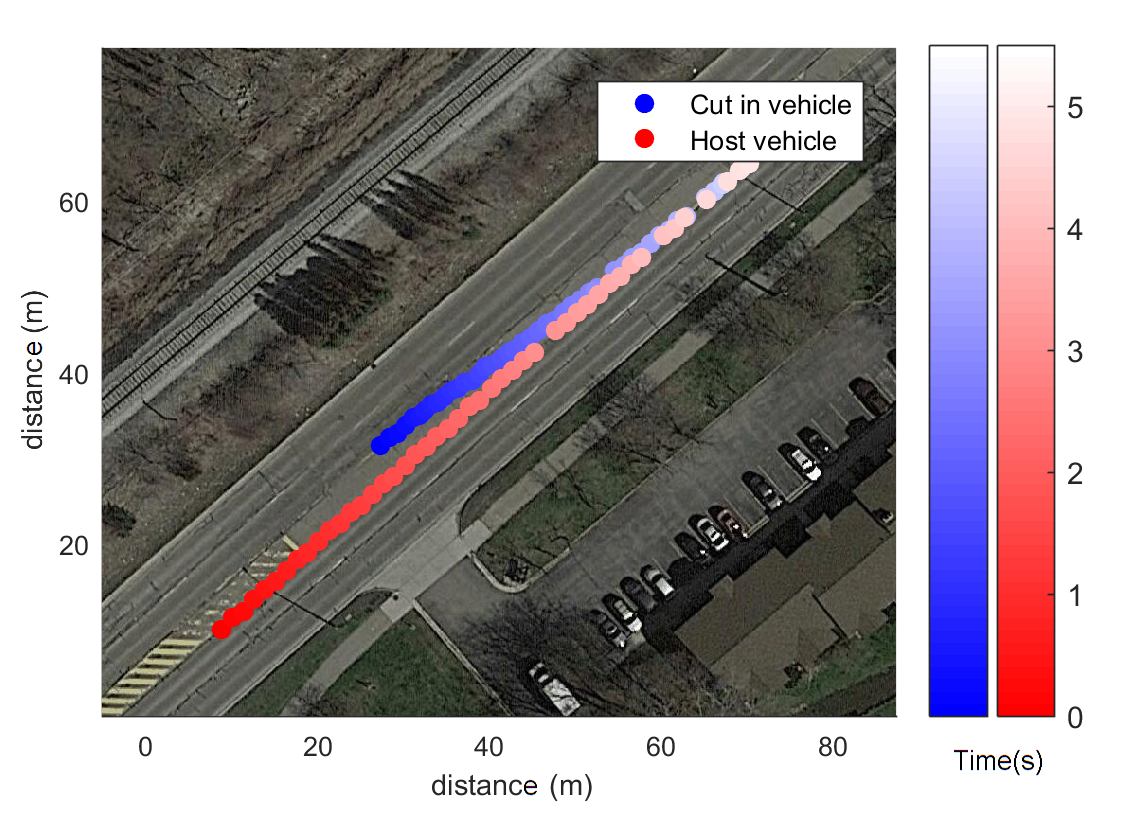}
\caption{Example of cut in events}
\label{fig:cutintb}
\end{figure}

\begin{figure}[h]
\centering
\includegraphics[width=0.75\columnwidth]{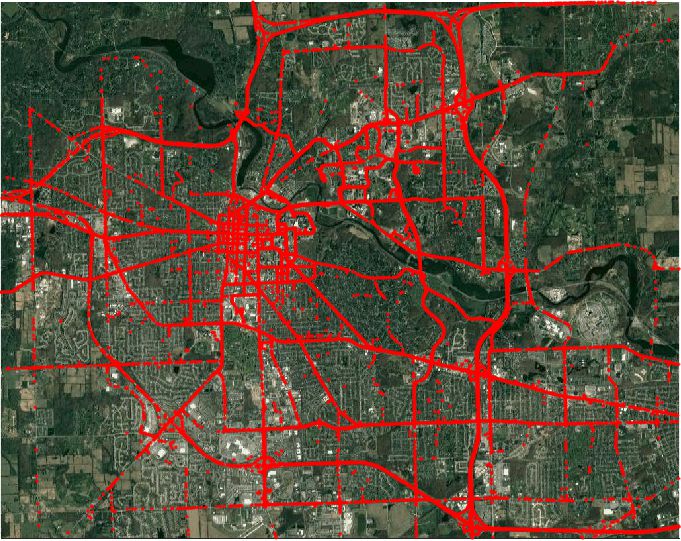}
\caption{Localization of cut in events}
\label{fig:cutintc}
\end{figure}

The cut in scenario records the events where a car cuts in the instrumented vehicle. Improper cut-in maneuvers would lead to crashes and pose challenges to the active safety systems~\cite{zhao2017accelerated,7989024}. In the TrafficNet database, the $Event$ table records the primary keys of the event, namely $Device$, $Trip$ and cut in time $c_t$ in seconds, while the $Sequence$ table records the data between $c_t-5$~s and $c_t +5$~s. When the instrumented vehicle drives along one lane and the $CIPV$ value of one car detected by the Mobileye changes from 0 to 1, then it means the front car moves to the lane of the instrumented vehicle. The algorithm to query cut in events is shown in Algorithm~\ref{alg:cutin}. An example is shown in Figure~\ref{fig:cutintb} and the distribution of the events is shown in Figure~\ref{fig:cutintc}.


 \begin{algorithm}[H]
 \caption{Cut in}
 \label{alg:cutin}
 \begin{algorithmic}[1]
 \renewcommand{\algorithmicrequire}{\textbf{Input:}}
 \renewcommand{\algorithmicensure}{\textbf{Output:}}
 \REQUIRE $DataFrontTargets$, $LaneChangeEvent$
 \ENSURE  $CutIn$
 \\ \textit{Initialisation} :
  \STATE $CutIn$ $\leftarrow \emptyset$,$cID \leftarrow 0$
  \FOR{each row $row_t$ in \\$DataFrontTargets\backslash LaneChangeEvent$}
  \IF{$CIPV_t=1$ and $CIPV_{t-1}=0 $\\$ObstacleId_{t}= ObstacleId_{t-1}$}
  \STATE $row_t \leftarrow cID$
  \STATE $CarFollowing \leftarrow row_t$
  \STATE $cID = cID+1$
  \ENDIF
  \STATE $t = t + 1$
  \ENDFOR
 \RETURN $CutIn$
 \end{algorithmic}
 \end{algorithm}

\subsection{Lane change}


In a lane change event, the vehicle changes from one lane to another. The lane change maneuver exposures the vehicle under the possibility of conflict and crash. Hence this data can be used for accelerating the evaluation of driving models.

The table \textit{LaneChangeEvent} contains the time stamp on when the vehicle changes its lane during a trip. It consists of 6 columns: \textit{Device}, \textit{Trip}, \textit{CrossTime}, \textit{ChangeDirection}, \textit{XTime1}, \textit{XTime2}. The tuple \textit{(Device, Trip, CrossTime)} is the primary key indicates a distinct lane changing event of a vehicle; \textit{CrossTime} is the time stamp of a lane changing event when center of the vehicle passing a lane; \textit{ChangeDirection} indicates which side the vehicle goes to; \textit{XTime1} and \textit{XTime2} are the time when the wheels near the target lane and the wheels on the other side cross the lane respectively, i.e., \textit{XTime1} is the time when the left wheels cross the lane during the vehicle goes to its left lane.

The algorithm to query lane change event consists of two parts, shown in Algorithm~\ref{alg:TwoWheelChange} and Algorithm~\ref{alg:WheelExistTime} .Since the Mobileye sensor may be blocked during detection, the lane information is not consistent and the updating of the left lane and right lane are not synchronized. In this case, to deduce the lane change event from the dataset, first we find a set of events in which the distances to the left lane and the right lane respectively change within 1~s, done in Algorithm~\ref{alg:TwoWheelChange}; then the exact time when wheels on each side moving to the target lane is deduced in Algorithm~\ref{alg:WheelExistTime}; finally, redundant events happening within 2~s are removed from it. For each trip \textit{(device,trip)} from table \textit{TripSummary}, first find the time when the distances to both lanes change dramatically, where we set the interval as 2~m to 4~m. \textit{LaneDisL} and \textit{LaneDisR} are the lateral distances to the left lane and right lane from the center of the car respectively; $t1$ and $t2$ are the time the $LaneDisL$ and $LaneDisR$ change more than 2~m but less than 4~m respectively; considering the discontinuity in the $DataLane$, if $\|t1-t2\|<1$ and the corresponding detection quality is high enough, $t1$ will be recorded as the $CrossTime$ of a lane change event. The term $XTime1$ and $XTime2$ are computed as follows: for the left lane change, $XTime1$ is the first time the left wheel crosses the left lane within 1~s before the $CrossTime$ where 0.91 is half the average width of sedans in meter. Equivalently, these two variables indicate the minimum time stamp the vehicle reaches the lane and the maximum time stamp the vehicle is still on the lane.


 \begin{algorithm}[H]
 \caption{TwoWheelChange}
 \label{alg:TwoWheelChange}
 \begin{algorithmic}[1]
 \renewcommand{\algorithmicrequire}{\textbf{Input:}}
 \renewcommand{\algorithmicensure}{\textbf{Output:}}
 \REQUIRE $TripSummary$, $Datalane$
 \ENSURE  $TwoWheelChange$
 \\ \textit{Initialisation} :
  \STATE $TwoWheelChange$ $\leftarrow \emptyset$
  \FOR{each $t$ when lane change }
  \IF{$\exists$ $t_2$ $\in$ [$t-10,t+10$] s.t. the other lane change at $t_2$}
   \STATE $TwoWheelChange$ $\leftarrow t$
  \ENDIF
  \ENDFOR
 \RETURN $TwoWheelChange$
 \end{algorithmic}
 \end{algorithm}

 \begin{algorithm}[H]
 \caption{WheelExistTime}
 \label{alg:WheelExistTime}
 \begin{algorithmic}[1]
 \renewcommand{\algorithmicrequire}{\textbf{Input:}}
 \renewcommand{\algorithmicensure}{\textbf{Output:}}
 \REQUIRE $LaneChangeEvent$, $CrossTime$ $t_c$
 \ENSURE  $XTime1$ , $XTime2$
  \\ \textit{Left lane change} :
  \FOR{all $t_l\in[t_c-1,t_c],t_r\in[t_c,t_c+1]$ \\     and $QualityLeft(t_l)>1$ $QualityRight(t_r)>1$}
  \STATE $XTime1$ $\leftarrow$ $min(\{t_l| LaneDisL(t_l)+0.91>0\})$
  \STATE $XTime2$ $\leftarrow$ $max(\{t_r| LaneDisR(t_r)-0.91<0\})$
  \ENDFOR
  \\ \textit{Right lane change} :
  \FOR{all $t_r\in[t_c-1,t_c],t_l\in[t_c,t_c+1]$ \\     and $QualityLeft(t_l)>1$ $QualityRight(t_r)>1$}
  \STATE $XTime1$ $\leftarrow$ $min(\{t_r| LaneDisR(t_r)-0.91<0\})$
  \STATE $XTime2$ $\leftarrow$ $max(\{t_l| LaneDisL(t_l)+0.91>0\})$
  \ENDFOR
 \RETURN $XTime1$ , $XTime2$
 \end{algorithmic}
 \end{algorithm}

\begin{figure}[h]
\centering
\includegraphics[width=\columnwidth]{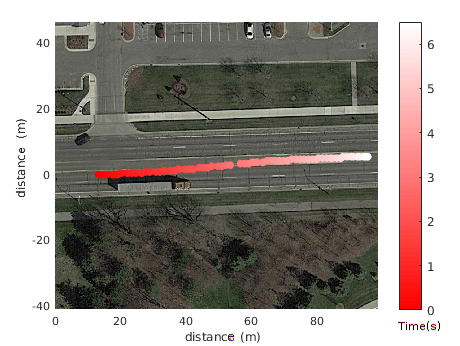}
\caption{Example of lane change events}
\label{fig:lanechangeb}
\end{figure}

\begin{figure}[h]
\centering
\includegraphics[width=0.75\columnwidth]{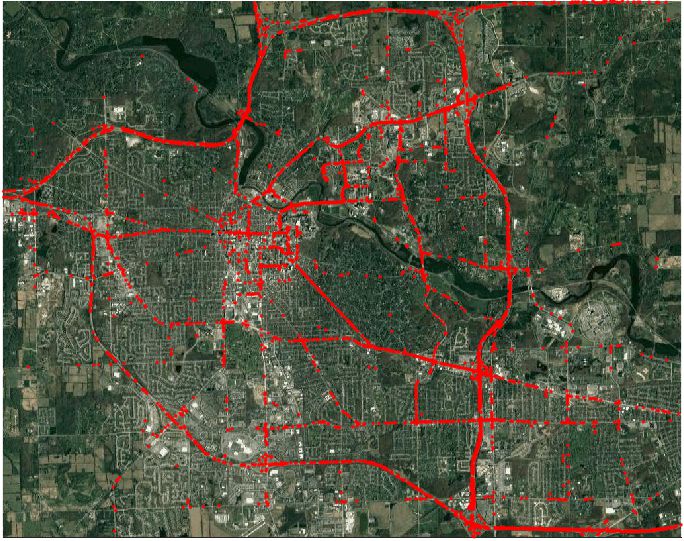}
\caption{Localization of lane change events}
\label{fig:lanechangec}
\end{figure}

\subsection{Pedestrian crossing}
$PedestrainCrossing$ logged the information on events when a pedestrian crosses the street in front of the vehicle ~\cite{chen2017evaluation} as shown in Fig.~\ref{fig:pedestriana}. In the TrafficNet database, We mainly use the range $D$ and lateral distance $L$ of the pedestrian related to the vehicle to deduce the trajectory of the pedestrian. To enable the analysis for the driving strategy, this dataset is provided in the time sequence of position and speed of the vehicle and pedestrian. It contains the position of the pedestrian in the vehicle coordinates and GPS information of the vehicle and pedestrian. The GPS data of pedestrian is deduced by transforming its position in the vehicle's frame to the global coordinates by the homogeneous transform. The algorithm to query this dataset is shown in Algorithm~\ref{alg:peds}. We query all the data in $DataFrontTargets$ where $TargetType=3$, which means the detected target is a pedestrian. A column named $pID$ is added for indexing. The false positive data are caused by the wrong detection and where the vehicle passed a pedestrian quickly, where a pedestrian appears less than 0.5~s.So we delete the events which have less than 5 rows.


 \begin{algorithm}[H]
 \caption{Pedestrian Crossing}
 \label{alg:peds}
 \begin{algorithmic}[1]
 \renewcommand{\algorithmicrequire}{\textbf{Input:}}
 \renewcommand{\algorithmicensure}{\textbf{Output:}}
 \REQUIRE $DataFrontTargets$, $DataWsu$
 \ENSURE  $Pedestrian$
 \\ \textit{Initialisation} :
  \STATE $Pedestrian \leftarrow \emptyset , pID \leftarrow 0$
    \\ \textit{Query data into $Pedestrian$} :
 \FOR{each row $r_t$ in $DataFrontTargets$}
 \IF{$TargetType=3$}
 \IF{$ObstacleID_t\neq ObstacleID_{t-1}$}
 \STATE $pID = pID+1$
 \ENDIF
 \STATE $r_t \leftarrow pID$
 \STATE $Pedestrian \leftarrow r_t$
 \ENDIF
 \ENDFOR
  \\ \textit{Delete false positive data} :
 \FOR{$count(pID)<5$}
 \STATE delete $r_{pID}$
 \ENDFOR
 \RETURN $Pedestrian$
 \end{algorithmic}
 \end{algorithm}


\begin{figure}[h]
\centering
\includegraphics[width=\columnwidth]{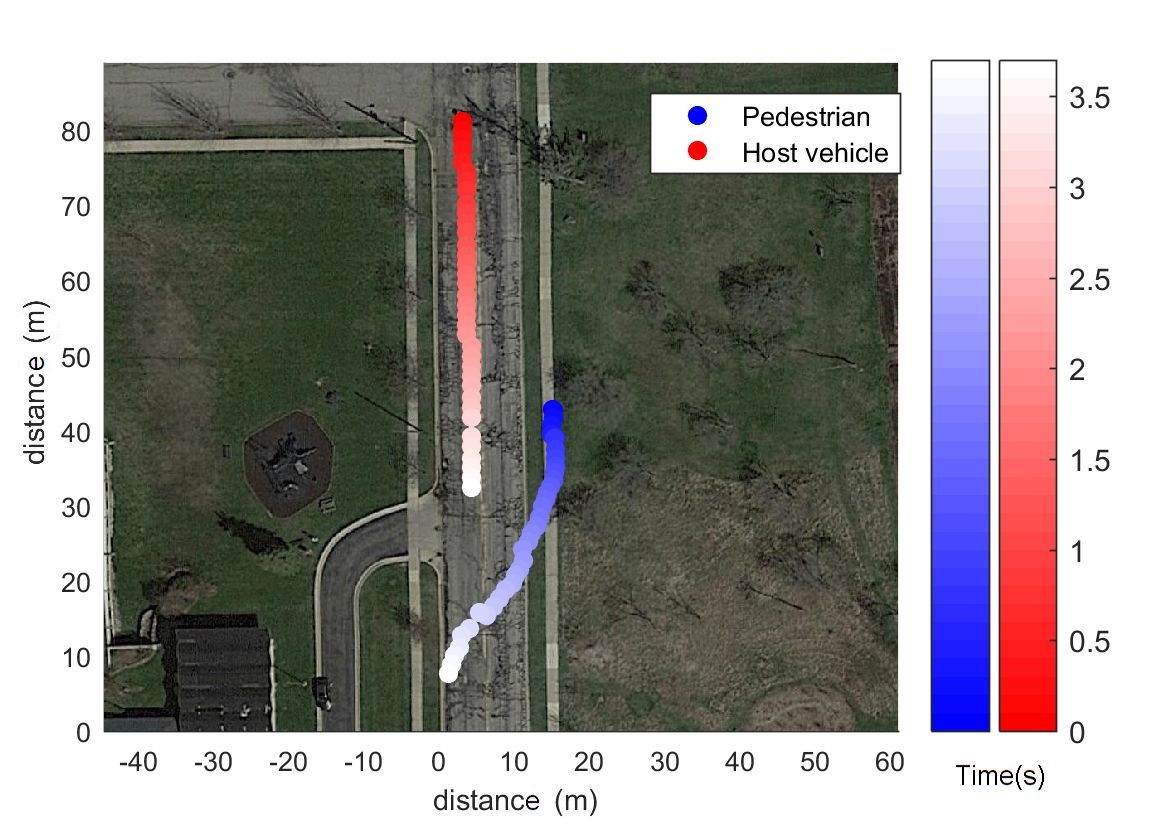}
\caption{Example of pedestrian}
\label{fig:pedestrianb}
\end{figure}

\begin{figure}[h]
\centering
\includegraphics[width=0.75\columnwidth]{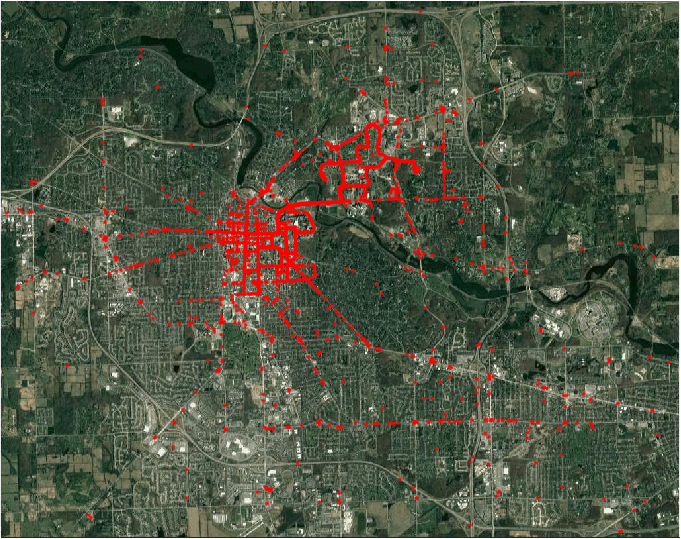}
\caption{Localization of pedestrian}
\label{fig:pedestrianc}
\end{figure}

\subsection{Cyclist}

The table $Cyclist$ records the positions, in terms of longitude and latitude, of a vehicle and a cyclist in time sequence when they encounter each other. The columns in this table includes: $Device$, $Trip$, $Time$, $StartTime$, $EndTime$. The variables $StartTime$ and $EndTime$ indicate the beginning when the Mobileye detects a cyclist and the time the Mobileye loses sight of the cyclist. Users can use the time stamps to query data on relative position and speed of the cyclist during each event, then get the knowledge of how the driver operates in different cases. The algorithm for querying $Cyclist$ is almost the same as querying $Pedestrian$, where the only difference is the $TargetType$ of cyclists is 4.

\begin{figure}[h]
\centering
\includegraphics[width=0.95\columnwidth]{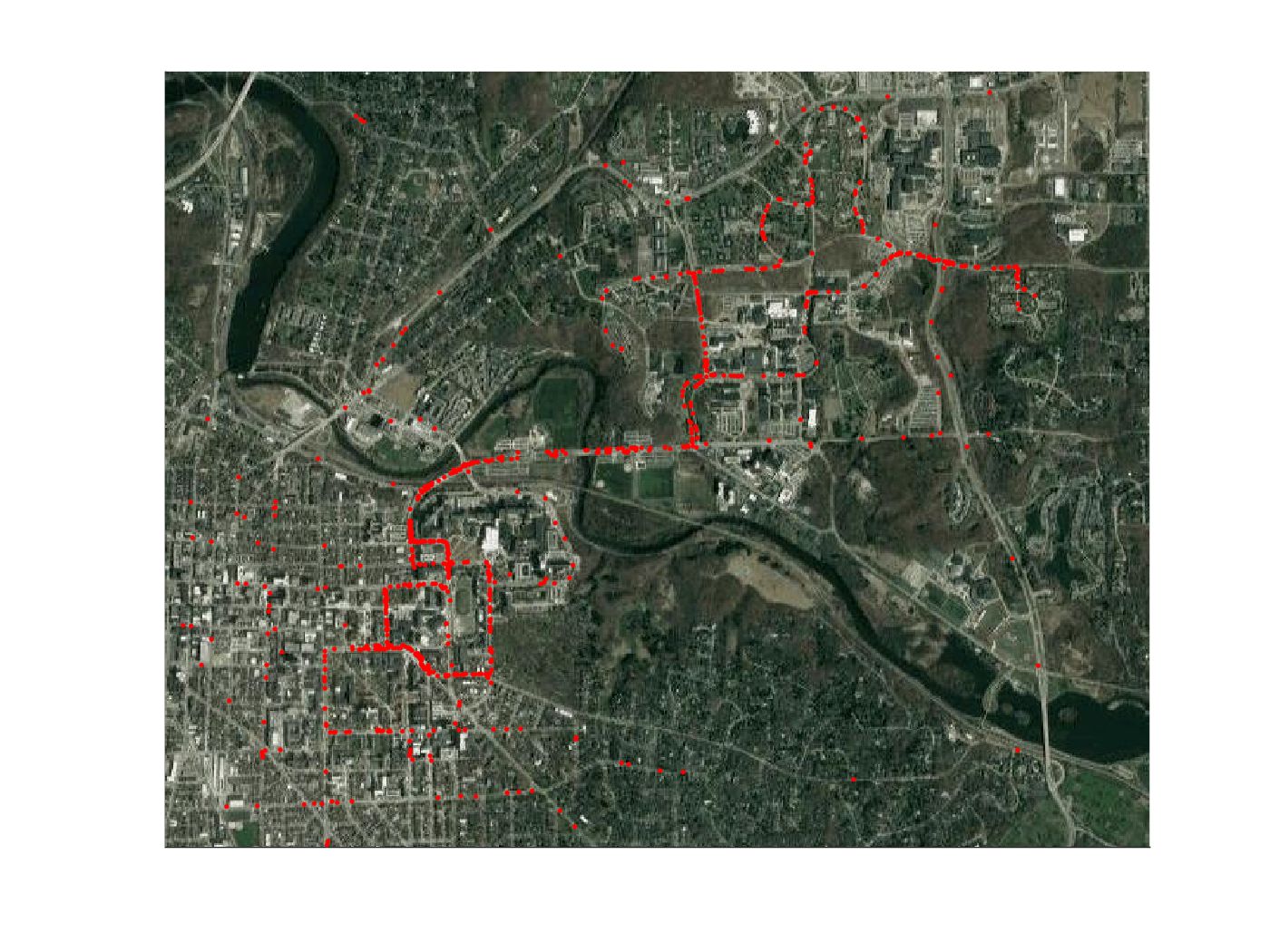}
\caption{Localization of cyclist}
\label{fig:cyclistc}
\end{figure}

The distribution of the cyclist events is shown in Fig.~\ref{fig:cyclistc}. This scenario is sparser than the others and mainly happens in the downtown area.

 \begin{algorithm}[H]
 \caption{Cyclist}
 \label{alg:cyclist}
 \begin{algorithmic}[1]
 \renewcommand{\algorithmicrequire}{\textbf{Input:}}
 \renewcommand{\algorithmicensure}{\textbf{Output:}}
 \REQUIRE $DataFrontTargets$, $DataWsu$
 \ENSURE  $Cyclist$
 \\ \textit{Initialisation} :
  \STATE $CyclistEvent \leftarrow \emptyset , cID \leftarrow 0$
  \\ \textit{Query data into $Cyclist$} :
 \FOR{each row $r_t$ in $DataFrontTargets$}
 \IF{$TargetType=4$}
 \IF{$ObstacleID_t\neq ObstacleID_{t-1}$}
 \STATE $cID = cID+1$
 \ENDIF
 \STATE $r_t \leftarrow cID$
 \STATE $CyclistEvent \leftarrow r_t$
 \ENDIF
 \ENDFOR
   \\ \textit{Delete false positive data} :
 \FOR{$count(pID)=1$}
 \STATE delete $r_{pID}$
 \ENDFOR
 \RETURN $Cyclist$
 \end{algorithmic}
 \end{algorithm}


\begin{figure}[h]
\centering
\includegraphics[width=\columnwidth]{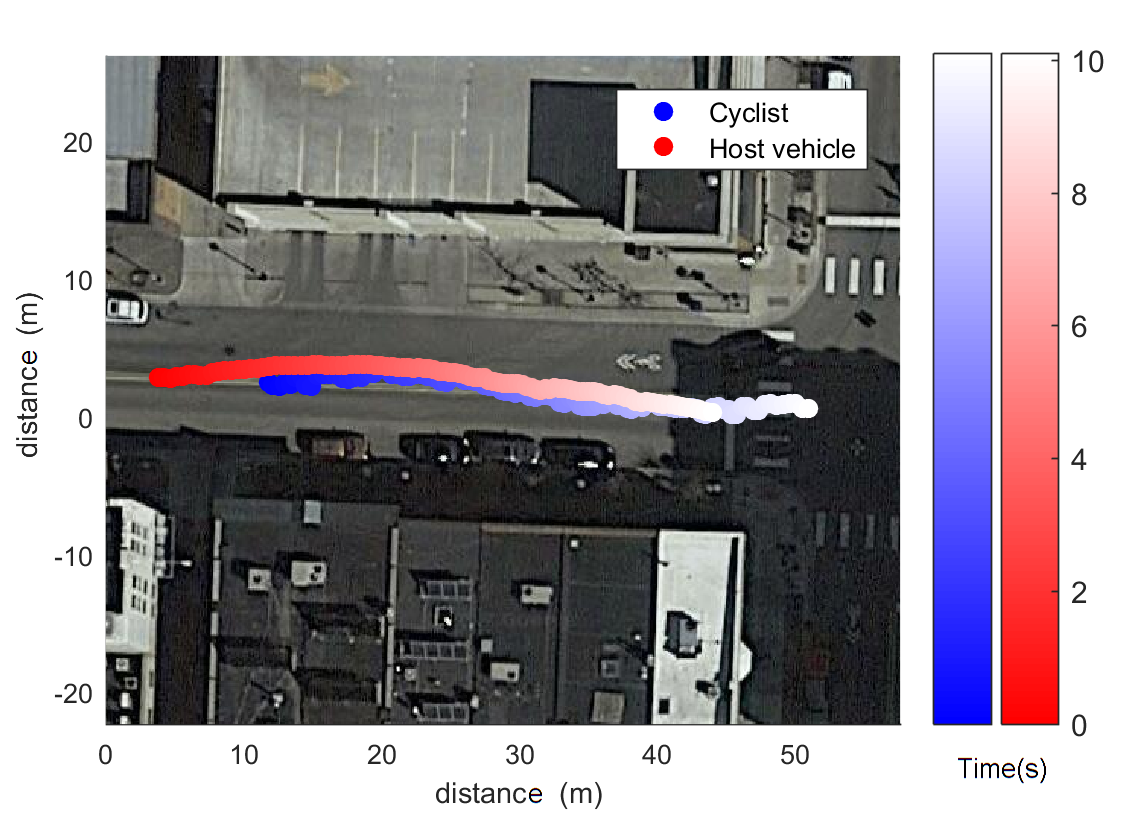}
\caption{Example of cyclist events}
\label{fig:cyclistb}
\end{figure}

\section{Conclusion}

In this research, We built the TrafficNet as a driving scenario-wise database extracted from the public data collected from the Safety Pilot Model Deployment program. Both the extracted events, raw data, and source code are provided on-line for free access. Compared to other database, TrafficNet provides scenario-based structure rather than chronological raw data, which may be directly used by automotive designers who are not familiar with database processing. Six types of scenarios were extracted, which, in total consist of 565,291 events. In the next step, we will develop more advanced scenario extraction tools and include more types scenarios.

\bibliographystyle{IEEEtran}
\bibliography{ref.bib}




\end{document}